# Surface-Driven Evolution of the Anomalous Hall Effect in Magnetic Topological Insulator MnBi$_2$Te$_4$ Thin Films

*Alessandro R. Mazza[1], Jason Lapano[1], Harry M. Meyer III[2], Christopher T. Nelson[1], Tyler Smith[1], Yun-Yi Pai[1], Kyle Noordhoek[1], Benjamin J. Lawrie[1], Timothy R. Charlton[3], Robert G. Moore[1], T. Zac Ward[1], Mao-Hua Du[1], Gyula Eres[1]#, Matthew Brahlek[1]\**

[1]Materials Science and Technology Division, Oak Ridge National Laboratory, Oak Ridge, TN, 37831, USA
[2]Chemical Sciences Division, Oak Ridge National Laboratory, Oak Ridge, TN, 37831, USA
[3]Neutron Scattering Division, Oak Ridge National Laboratory, Oak Ridge, TN, 37831, USA
Correspondence should be addressed to *brahlekm@ornl.gov, eresg@ornl.gov#

**Abstract**: Understanding the effects of interfacial modification to the functional properties of magnetic topological insulator thin films is crucial for developing novel technological applications from spintronics to quantum computing. Here, we report that a large electronic and magnetic response is induced in the intrinsic magnetic topological insulator MnBi$_2$Te$_4$ by controlling the propagation of surface oxidation. We show that the formation of the surface oxide layer is confined to the top 1-2 unit cells but drives large changes in the overall magnetic response. Specifically, we observe a dramatic reversal of the sign of the anomalous Hall effect driven by finite thickness magnetism, which indicates that the film splits into distinct magnetic layers each with a unique electronic signature. These data reveal a delicate dependence of the overall magnetic and electronic response of MnBi$_2$Te$_4$ on the stoichiometry of the top layers. Our study suggests that perturbations resulting from surface oxidation may play a non-trivial role in the stabilization of the quantum anomalous Hall effect in this system and that understanding targeted modifications to the surface may open new routes for engineering novel topological and magnetic responses in this fascinating material.

# 1. Introduction

The topological description of materials' properties has led to the discovery of many novel states of matter—each of which demonstrate vast applicability in real-world devices.[1] This field has rapidly expanded over the past decade due to the emergence of 1-dimensional (1D)[2–4] and 2-dimensional (2D)[5–7] topological surface states which exhibit linear, Dirac-like dispersion and near perfect spin-momentum locking.[8,9] When combined with ferromagnetism, the degeneracy at the Dirac point is lifted by the violation of time-reversal symmetry, which gives rise to chiral, 1D quantized edge states in the quantum anomalous Hall effect (QAHE).[10] This was initially achieved in magnetically doped tetradymite topological insulator (TIs) systems, in particular Cr/V-doped $(Bi_{1-x}Sb_x)_2Te_3$.[11–14] Recently, the $MnBi_2Te_4$ system has arisen as a new paradigm where intrinsic magnetism arises due to the large moment of the Mn atoms, while maintaining Bi-Te band inversion necessary to host topological surface states.[15] In comparison to magnetically doped TIs, the high density and large $Mn^{2+}$ moment could open a much larger surface gap[15], which is critical for boosting the QAHE temperatures to as high as 77 K (liquid nitrogen) and above.[16] The tetradymite 5-atom $Bi_2Te_3$ quintuple layered (QL) system is modified crystallographically by the intercalation of a MnTe layer in the center of the QL creating a septuple layer (SL).[17] Together, the $Bi_2Te_3$ and $MnBi_2Te_4$ layers form the basic units for the homologous series $MnBi_2Te_4(Bi_2Te_3)_n$ where $MnBi_2Te_4$ is spaced by $n$ layers of $Bi_2Te_3$. This system has been successfully synthesized both by bulk[18–21] growth techniques as well as thin films by molecular beam epitaxy (MBE).[22–25]

Several recent studies have highlighted the intriguing interplay between topology and the novel A-type antiferromagnetism (AFM) in this system, where the Mn spins align ferromagnetically in-plane, which then couple antiferromagnetically between the layers.[19,26] This ordering enables tuning of the magnetic ground state through layer thickness. For an even number of layers, the net moment vanishes via exact compensation, whereas an odd number of layers result in a net moment, thereby breaking time-reversal symmetry. This has guided the initial observation of the QAHE in an exfoliated bulk crystal as well as an analogous effect of an axionic insulator in cases where time-reversal symmetry is preserved.[27–29] Despite this success, many questions remain. Central to ongoing research is the question of the role that defects play in stabilizing the magnetic ground state. In comparison with the binary tetradymites, defects are far more abundant in $MnBi_2Te_4$[14,16,30,31] because of the more complex chemistry.[32] The defects and disorder are likely major factors in the controversy related to the presence or absence of surface gaps in photoemission, and the diverse character of the reports needed to achieve quantization with questions on how to realize reproducibility.[33] The novel layered magnetic and crystal structures open many routes to modify the topological state and global magnetic phase by targeted changes to the surface, for example, through oxidation,[34,35] which, is critical to fully understand the factors that give rise to the QAHE. There are two recent reports suggesting that inadvertent ambient oxidation may play an important role in the manifestation of the QAHE.[27,36] The first work reports that oxidation starts at and proceeds from step edges and not from the basal planes, a result that can have important implications for the stability and behavior of the edge states.[36] The second work concerns the use of an $Al_2O_3$-exfoliation technique used in fabrication of the sample for the only report of zero-magnetic field QAHE in $MnBi_2Te_4$.[27] Although, the role of inadvertent oxidation was not addressed in this work, considering the delicate nature of topological states a closer look is necessary before it can be ruled out. Direct studies of the surface of $MnBi_2Te_4$ are limited and connecting the surface chemistry to the QAHE may lend insight to the reproducibility and conflicting results reported in the literature.

Here we combine systematic measurements with first principles calculations of the electronic and magnetic evolution of MBE-grown finite thickness $MnBi_2Te_4$ thin films during post growth oxidation to provide new insight into the formation of the oxide layer and subsequent large changes to the electronic and magnetic structure. We used conventional x-ray photoemission spectroscopy (XPS) in ultrahigh vacuum (UHV) to determine the chemical composition changes resulting from air induced oxidation as a function of long-time exposure. To exclude water that is known to promote $Bi_2Te_3$ oxidation we compare these results with *real-time* measurements in a low-pressure oxygen environment using near ambient pressure (NAP) XPS.[37] Because XPS has a finite probing depth we used neutron reflectometry (NR) to determine the thickness resolved evolution of the oxide layer that forms in $MnBi_2Te_4$. The oxide layer formation is found to be confined to the top 1-2 SL. Transport measurements, which are sensitive to bulk properties of the films, show a similarly strong evolution of the electronic properties (resistivity and carrier density) but also show a dramatic change to the finite thickness magnetic character. First principles calculations reveal that the formation of the oxide layer is energetically favorable, consistent with our experimental results. Importantly, these calculations are used to further elucidate the effects on the surface state properties, where prominent metallic, in-gap states are found to emerge. The role of these results, particularly in understanding the conflicting observations in bulk crystals, is crucial going forward to understand the conditions for the QAHE in this series of compounds and, more broadly, to design and utilize the novel electronic and magnetic properties in $MnBi_2Te_4$.

## 2. Results and Discussion

The chemical composition changes induced by oxidation and the evolution of the electronic structure produced by post growth air exposure of the $MnBi_2Te_4$ films was determined by XPS measurements in UHV using a monochromated Al $K_\alpha$ X-ray source. The XPS measurements were performed on high-quality MBE-grown films described in Ref.[25], which were transferred to the XPS chamber in air with an exposure time of ~2 minutes. The survey spectra for the as grown film and the same film after 21.5 hours of air exposure are compared in the Supporting Information. In addition to the clear increase of the O 1s peak intensity these spectra show the Bi 4f, Te 3d and Mn 2p core levels, for which high resolution scans as a function of increasing air exposure time are given in Figure 1(a). The Bi 4f core level is split by spin-orbit interaction into a characteristic doublet, $4f_{5/2}$ and $4f_{7/2}$ at 162.9 eV and 157.6 eV binding energy (BE) separated by 5.3 eV.[34,38] The Bi-O peaks are separated from Bi-Te peaks by about 1 eV and show up on the high BE side of Bi-Te peaks. The Bi-O peaks at 164.1 eV and 158.9 eV BE, except perhaps as a weak shoulder are not present in the as grown films, and after 1290 minutes of air exposure the Bi-Te peaks are only present as a weak shoulder on the low BE side of the Bi-O peaks. Similarly, the Te 3d core level consists of two spin orbit components $3d_{3/2}$ and $3d_{5/2}$ at 582.8 eV and 572.5 eV BE separated by 10.3 eV.[34,38] The Te-O peaks separated by about 3.3 eV show up on the high BE side of the Mn/Bi-Te peaks. The Te-O peaks at 586.2 eV and 575.8 eV BE are clearly observable as weak peaks already present in the as grown films, and after 1290 mins of oxidation the Mn/Bi-Te peaks are still clearly present. It is important to note that no shift of the Bi and Te core levels from the as grown position was observed, indicating that there is negligible doping of the $MnBi_2Te_4$ film.[34]

The Mn 2p core level also consists of the characteristic spin-orbit doublet $2p_{1/2}$ and $2p_{3/2}$ at 652.8 eV and 641.3 eV BE separated by 11.5 eV. The main component of the Mn 2p peaks correspond to the $Mn^{2+}$ oxidation state that matches both MnTe and MnO at BE of 640.8 eV.[39] The Mn 2p peaks are difficult to fit because in addition to ambiguous separation of the MnTe and MnO components these peaks have additional satellite components that are related to ligand core-hole interactions in the photoemission process.[39] Oxidation of Mn typically produces $MnO_2$, which would show up at 642.1 eV BE. Fitting the Mn 2p with these components can produce arbitrary results and instead of fitting we directly compare the Mn 2p spectra for the as grown film and long-time air oxidation in Figure 1(a) where the data have been normalized for comparison. The close overlap of these two spectra indicates that 21.5 hour long oxidation produces no changes of the Mn layer detectable by XPS.

The plots of the intensities as the ratio of the Bi-O and the Te-O peaks and the total intensity as a function of air exposure time reveal that oxidation of Bi and Te do not occur at the same time nor at the same rate. While the onset of Bi oxidation is immediate, Te oxidation accelerates only after a substantial, more than 10-hour long induction, period. These data agree with findings of a thermodynamic study that Bi is more reactive than Te, and that a Te terminated surface can be completely inert under some circumstances.[40] Another intriguing observation related to the individual elements, Bi and Te, oxidation behavior are the data obtained by *real-time* NAP XPS, which was performed at an $O_2$ pressure of 1.5 Torr and is shown in Figure 1(b). The time used for the oxygen exposed sample is scaled by the relative oxygen partial pressure to be compared with the air exposed samples according to *time*$^*=P_{O2,\text{XPS}}/P_{O2,air}\times time$, where $P_{O2,\text{XPS}} \approx 1.5$ Torr and $P_{O2,air} \approx 159$ Torr. The scaled data reveal that oxide formation in the pure oxygen exposed sample is fast and reaches about 40% already at *time*$^* \approx 10$ minutes. In contrast to the air exposed samples the rate of change is found to be nominally the same for Bi and Te in the NAP XPS data. The difference in the oxidation rates in air and oxygen is likely attributed to the presence of $H_2O$ that was found to promote oxidation of $Bi_2Te_3$ in air.[37] It is intriguing that the fast oxidation rates in our experiments disagree with the results of similar NAP XPS measurements that found negligible surface reactivity at somewhat lower oxygen pressures up to 75 mTorr.[37] It is possible that a 2 minute air exposure is already sufficient to trigger the oxidation process, suggested by the presence of the Te-O peaks in the XPS spectrum of the as grown film. This discrepancy in the behavior and rates of $MnBi_2Te_4$ oxidation is consistent with current literature reporting contradicting results about the oxidation process of this family of compounds under nominally similar conditions.

To understand the chemical stability of oxygen in the $MnBi_2Te_4$ SL structure, we performed first principles calculations (see Supporting Information for details). Despite similar valence, the large size mismatch and slow kinetics at room temperature make it unlikely that oxygen will replace Te directly in $MnBi_2Te_4$ at ambient conditions. Therefore, it is more likely that oxygen is interstitially bonded within the structure. Finding the lowest energy positions for oxygen within the SL was done by calculating the formation energies, which are shown in Figure 1(c) for four possible configurations. Here, the oxygen atoms were allowed to relax to the minimum energy position within the SL and ambient conditions were assumed ($T = 300$ K and $P_{O2} = 200$ Torr). The formation energies are found to be negative for all positions, which indicates that oxygen molecules on the surface favor dissociating into atomic oxygen that is stabilized within the SL structure. Moreover, the interstitial sites closest to the surface (Figure 1(c) i-ii) are found to be higher in energy relative to the sites near the inner most Te layer (Figure 1(c) iii). This implies that an oxygen is likely to chemisorb onto the surface, and after dissociating diffuse into the SL structure. Further, the site nearest the Mn layer (Figure 1(c) iv) is found to be higher energy than outer sites, and of significantly higher energy than the temperature scale. This large energy difference implies that the Mn layer will likely act as a kinetic barrier to prevent further oxidation. Overall, the lowest energy position for oxygen is found to be at position iii (-0.729 eV per oxygen) when bonded into the SL, and the structure is allowed to relax. The relaxed structure is shown in Figure 1(d). Here, the oxygen sits between the Bi and Te site that is closest to the central Mn layer. The stability of the oxygen on this site relative to the site at position ii (-0.448 eV per oxygen) may be related to the resulting distortion, which is predominantly out-of-plane for iii and in-plane for ii. Importantly, oxygen at site iii is found to slightly perturb the Mn-Te bonding, which, as discussed below, likely has implications on the Mn-Mn magnetic ordering.

The difference in the elemental oxidation process between Bi and Te observed in the time dependent air exposure XPS data leads to compositional nonuniformity in the films. The oxidation rate difference is found to be related to the respective bond strengths, with the Bi-Te being the strongest bond (-1.48 eV) followed by Bi-O (~ -1 eV) and Te-O (-0.76 eV).[40] The strength of this bond indicates that Bi reacts faster with O than with Te, which gives rise to complex compositionally nonuniform atomic transport processes. A recent scanning tunneling microscopy (STM) study found that perfect Te surface termination created by exact cleavage between the Te-Te planes is highly stable compared to the Bi terminated surfaces that are prone to faster oxidation.[41] The oxidation process starts by dissociation of $O_2$ initially chemisorbed on the surface to create highly reactive O species. It was reported that the Te termination compared to Bi termination considerably impedes the inward diffusion of O into the lattice creating a spatial nonuniformity in the oxidation process. This slower element-sensitive oxidation component combined with the already mentioned faster reaction component that occurs at the step edges reinforces the spatial nonuniformity of the oxidation process. The compositional and spatial nonuniformity produced by the oxidation may have profound significance for the electrical and magnetic transport properties of the $MnBi_2Te_4$ films, as discussed below.

In contrast to XPS that is highly surface sensitive but has a finite penetration length, NR measurements are ideally suited to resolve the oxygen density profile across the entire sample thickness.[42,43] The changes in the intensity of the Bi and Te peaks with oxidation time can be used to estimate the oxide film thickness by the Hill equation.[34,38,44] However, the finite penetration depth and the estimates of the electron inelastic mean free paths for the Hill equation introduce large uncertainties that obscure accurate determination of the oxide layer thickness by fixed energy X-ray source XPS. Instead, we use NR to determine the oxide layer thickness. Figure 2(a) shows reflectivity data and resulting fits for a 7 SL $MnBi_2Te_4$ film. This film was exposed to air on the timescale of hours and then stored in a desiccator during transfer from the MBE to the beamline prior to being mounted in the cryostat and pumped down. Therefore, these results provide a picture of the oxide layer in the long-time, stable regime.

Here, the detailed decay with increasing scattering vector ($q_Z$) and the character of the oscillations are fingerprints of both the geometry (thickness and roughness/gradients), and the atomic constituents. These characteristics can be quantified by a detailed model[45] and the resulting fit is plotted as a solid curve on top of the data in Figure 2(a), which is in excellent agreement with the data (see Supporting Information for more detail). The scattering length density (SLD) profile, $\rho_{SLD}$, extracted from the fit to the data is plotted in Figure 2(b) as a function of depth relative to the substrate/film interface. The SLD is highest in the $Al_2O_3$ substrate because of the large value of the SLD for oxygen and its dense packing. It falls sharply within the $MnBi_2Te_4$ films because of the negative SLD for Mn. The jump and the peak in the $MnBi_2Te_4$ SLD near the film surface is a clear indication of oxygen incorporation leading to formation of the oxide layer. We note that the NR results only reveal the large SLD of the surface with no chemical specificity, and other oxygen rich species in the air, carbon, or nitrogen would similarly raise the SLD of the surface if intercalated. However, combined with the results of the XPS measurements we assume that the increase in the SLD is purely caused by oxygen. These fits reveal that oxidation is tightly confined to ~ 2 nm ± 0.5 nm of the $MnBi_2Te_4$ thin film surface. Assuming a single SL, this corresponds to an oxygen content of $MnBi_2Te_4O_{6.5\pm0.5}$, which is overall consistent with the XPS analysis. As a complement, bulk sensitive x-ray diffraction shown in Figure 2(c) was performed. The well-resolved 00L peaks show that the crystal structure is preserved. A scanning transmission electron microscopy (STEM) image taken on a NION UltraSTEM at 200kV is shown in the inset of Figure 2(c). The sample was prepared by focused ion beam liftout followed by low energy (900eV) Ar ion milling in a Fischion NanoMill. This similarly highlights that the SL structure of the films is preserved after long exposure to air. Together these data indicate that other than the top layers which oxidize as a function of exposure time the MnBi$_2$Te$_4$ is quite air stable, as schematically illustrated in Figure 2(d).

The magnetic response in A-type AFM materials such as MnBi$_2$Te$_4$ depends strongly on the thickness. For the ideal case, the net moment is compensated for an even number of layers and for odd numbers of layers there is a residual non-compensated moment.[46,47] Real materials show a richer response where both even and odd layer number thin films can exhibit non-zero magnetic moment.[48] As such, transport measurements are an extremely sensitive tool for identification of surface-driven effects on these magnetic and electronic properties. Compiled in Figures 3-5 are transport measurements performed on a nominally 6 SL sample (see Supporting Information for x-ray reflectivity) that was measured with an initial exposure of about 10 minutes to over 5000 minutes. Measurements were performed using pressed indium contacts in van der Pauw geometry down to a base temperature of 2 K. Magnetic field dependent resistivity data (applied out-of-plane) was symmetrized and the Hall resistivity was anti-symmetrized to remove a small amount of intermixing of the other components.

To understand the evolution of the bulk AFM, magnetoresistance measurements, MR=($R(H)$-$R(H=0T)$)/$R(H=0T)$×100%, where $R$ is the sheet resistance and $H$ is the applied magnetic field, are shown in Figure 3(a) for temperature $T$ = 2 K and exposure times from 10 to 5820 minutes. These data show two clear features: The first feature is the prominent butterfly-like hysteresis that occurs near zero field, reported in the literature for both MnBi$_2$Te$_4$ flakes and thin films[25,29], which is indicative of the finite thickness ferromagnetism. The magnitude of the hysteresis is found to increase with exposure up to 60 minutes, and then reduce after 1860 minutes. Moreover, the magnetic field scale where the loops close is pushed to higher fields, which can be seen in the zoom-in in Figure 3(e). The former behavior suggests the ferromagnetic domain formation responsible for magnetoresistance changes, which may be due to anisotropy changing or pinning of domain boundaries. The more complex loop shape and critical fields where the loops close is more apparent and parallels the changes in the anomalous Hall effect, which will be discussed in depth later. The second feature is the peak at around $|H_{SF}|$ ≈ 3-4 T, which is marked by vertical dashed arrows. This is a well-known signature of the spin-flop transition, where the AFM anisotropy direction rotates to be orthogonal to the direction of the applied magnetic field and eventually saturates at high-fields, as schematically shown in Figure 3(b). The field where this peak occurs was empirically found to depend closely on the thickness in thin films[48]. This thickness dependence can be easily understood since the net energy to rotate the anisotropy direction of an AFM depends on both surface and bulk magnetic contributions. For A-type AFMs such as MnBi$_2$Te$_4$, this translates to the material becoming more ferromagnetic-like (i.e. softer) in the thin regime. Tracking $H_{SF}$ versus both temperature and time reveals several important points. First, the spin-flop fields versus temperature are shown in Figure 3(c) where $H_{SF}$ monotonically decreases and vanishes at the bulk Neel temperature, around 25 K, and is found to qualitatively follow a Bloch-type law. This indicates that overall, the bulk-like AFM is preserved for all times. However, the magnitude of $H_{SF}$ is found to reduce with increasing time. This can be seen clearly in Figure 3(d), where $H_{SF}$ is plotted versus exposure time for $T$ = 2 and 15 K, where both temperatures show a similar trend. This reduction is consistent with the sample becoming magnetically thinner, or analogously the formation of a secondary magnetic layer, related to oxide formation. However, the change is more intricate than this simple assertion. This complexity can be seen by examining the curve shape of the normalized MR for $H<H_{SF}$, as shown in Figure 3(e) where a hump forms after long exposure times. The smaller $H_{SF}$ of ~ 2.4 T is consistent with a secondary weaker spin-flop of a thin layer, which emerges at very short times, steadily grows, and then saturates at longer times. The emergence of this feature coincides with the timescale at which full oxidation is achieved.

Hall effect measurements were performed to examine and quantify how the ferromagnetism observed in the MR at low magnetic fields changes with increasing surface oxidation. These results, $R_{xy}$ versus $H$, are shown in Figure 4(a). These data reveal several prominent features in the low-, mid-, and high-field regimes. First, in the low-field regime ($H < 2$ T) there is the aforementioned hysteresis due to ferromagnetism. Examining the low-field hysteresis gives insight into the changing electronic and magnetic character and is discussed below. In the mid-field range ($H \approx 3$ T) there is a kink due to the spin-flop transition. In the high magnetic field regime ($H > 7$ T) the curve flattens and is nominally linear in $H$ due to the free electron response. As shown in Figure 4(b), the high-field slope gives the carrier density versus time, which is plotted alongside the carrier density extracted at room temperature. Comparing these two temperatures eliminates confounding of magnetic contributions from the Hall effect. These data show that the carrier density is quite low, $1\text{-}2\times10^{13}$ cm$^{-2}$ and relatively independent of time on the order of hundreds of minutes. However, at the longest times, it is found that the dominant carrier type changes from n-type to p-type, which is quite unexpected considering the absence of core level shifting in the XPS spectra. This result suggests the following: (1) The incorporation of oxygen occurring is not a simple atomic substitution, but an electron acceptor behavior related to Bi-O-Te bond formation in oxidized MnBi$_2$Te$_4$. (2) The Hall effect probes the bulk of the film, which implies that there is likely a secondary in-diffusion of oxygen or other atmospheric gas into the bulk of the film; the density of this is small in comparison to the surface oxides since this is not apparent in the neutron reflectivity, yet transport properties are extremely sensitive to small changes, as has been well-established for Bi$_2$Se$_3$.[35,49] Overall, this behavior emphasizes the complexity of both the ambipolar defects possible in this class of compounds as well as the nature of atmospheric dopants.

Examining the low-field hysteresis shows an equally dramatic evolution where the sign of the anomalous Hall contribution changes between 60 and 1860 minutes. This can be seen in Figure 4(c), where $R_{xy}$ at zero-field is plotted versus time (taken on the up-sweep, i.e. $H \to$ -9 T to 9 T). This data further reveals a clear overall trend where $R_{xy}$ linearly increases with time until ~ 100 minutes and then begins to decrease and eventually changes sign at ~ 300 minutes. The initial rise can be understood as an effect of increased scattering. For metals with resistivity ($\rho$) $> 10^{-4}$ $\Omega$cm, i.e. semimetals, degenerately doped semiconductors, and highly disordered metals, the anomalous Hall resistance is known to scale with the resistance ($R_{xy}$ ~ $R_{xx}^p$, $p>1$).[50] This is consistent with the observed change in resistivity, shown in Figure 4(d), where both the room temperature and low temperature resistivities are found to increase with increasing time up to ~ 100 minutes. In contrast, the change in sign signals a more complex evolution of both electronic and magnetic properties since the sign of the anomalous Hall effect convolutes both the electronic and magnetic structures. It should be noted that this sign reversal is not simply related to the n- to p-type transition.[50] The origins can be unraveled by looking more deeply into the character of the loop shapes.

For finite thickness MnBi$_2$Te$_4$ it has been empirically found that the hysteretic loops of the anomalous Hall effect can be decomposed into multiple individual loops.[48] Specifically, this can be quantified by fitting the data to a function of the form

$$R_{xy,A}(H) = \sum_i R_{xy,Fi} \tanh w_i (H - H_{C,i})$$

Where $i$ indexes the individual anomalous Hall component, which is found to be either 2 or 3 in this work, $R_{xy,Fi}$ is the magnitude of the response, $w_i$ is the width of the transition, and $H_{C,i}$ is the coercive field (where the anomalous Hall component vanishes). The hyperbolic tangent function is derived from a Langevin-type analysis for the net magnetization in a magnetic field.[51] As the understanding of this effect in $MnBi_2Te_4$ is still evolving, we chose to follow the interpretation presented in Ref.[48] where one component was interpreted to be due to the finite-thickness-induced, non-compensated moment of $MnBi_2Te_4$ which persists up to the Neel temperature of 25 K. The sign of this component is found to be ambipolar, positive for even-layer samples and negative for odd-layer samples with a larger $H_C$ (~1-2 T for 6 SL). The second component, hypothesized to be an impurity, is found to vanish below 15 K, always have negative $R_{xy,F}$, and to have a significantly smaller $H_C$ (~0.2-0.3 T). We fit the Hall effect data to this model (adding an additional polynomial, $H$ and $H^3$, background term to account for the free electron response), with the results shown in Figure 5 for temperature and for various exposure times. Qualitatively, the data at 2 K for exposures of less than 30-40 minutes is composed of two loops, $R_{xy,F1}$ that is positive and $R_{xy,F2}$ which is negative. This can be seen in the raw data and fits (Figure 5(a-d), i), and from the decomposed terms in (Figure 5(a-d) ii-iv). In going from low to high temperatures >15 K, this two-loop shape gives way to a single loop, which is consistent with the positive term being the non-compensated contribution and the negative component being the impurity-like contribution.[48] The sign and magnitude (blue curve in Figure 5(e)) of the non-compensated component is consistent with the even SL layer thickness of this sample while the negative component is consistent with the impurity-like term (black curve in Figure 5(e)). For comparison we have included estimates at $t = 0$ minutes from Ref [48] for the Te capped 6 SL sample indicated as blue dashed arrows (surface term) and black dotted arrows (impurity term).

Following these components with increasing exposure highlights a very interesting evolution of the electronic and magnetic structures. First, at short times ($t = 10$-40 minutes, as well as the Te capped samples from Ref. [48]) it is seen that the magnitude of the impurity-like term, $R_{xy,F2}$, steadily increases. Similarly, the surface term also increases, albeit less dramatically. For Te capped samples $R_{xy,F}$ increases for reduced thickness[48], and suggests the sample's effective magnetic thicknesses becomes thinner, agreeing with the above results of the MR. $H_C$ is nominally the same for both components and exhibits a small increase for times up to 40 minutes. Surprisingly, at around 40-60 minutes there emerges an additional component with negative amplitude, $R_{xy,F3}$. Within this transition regime, the shaded region in Figure 5(e), the fits to both 2-component and 3-component models are poor. For 60 minutes and beyond, the 3-component model fits the data well, as shown in Figure 5(b-d). Connecting the components before 30 and after 60 minutes presents a challenge, and to indicate this ambiguity we chose not to connect the data points across this regime. Using the criteria of the ambipolar sign of $R_{xy,F}$, large $H_C$ (>0.5 T), and non-zero magnitude at 15 K, we, therefore, take the components $R_{xy,F1}$ and $R_{xy,F3}$ to be non-compensated-like terms and $R_{xy,F2}$ to be the impurity-like term since it is always negative with a relatively small $H_C$. Interestingly, the non-compensated terms ($R_{xy,F1}$ and $R_{xy,F3}$) eventually change sign at longer times, whereas the impurity-like term ($R_{xy,F2}$) is nominally unchanged, indicating the properties giving rise to this term are nominally constant with exposure time. Consistent with the analysis of the MR (Figure 3), this together implies that the effective thickness, derived from the non-compensated bulk AFM, is reduced, or perhaps the film is magnetically split into portions of even and odd thicknesses.

It is interesting to explore whether these independent magnetic regions are through the thickness or lateral, both of which are consistent with the structural and chemical analysis above of a 1-2 nm thick surface oxide layer and a bulk structure that is unaffected by oxidation. The first scenario would have the near surface layer being magnetically split from the remaining bottom portion of the film, driven possibly by exchange striction to the magnetic surface oxide (see below); this is consistent with the large magnitude of $R_{xy,F}$ which is similar to several SL thick capped samples[48]. Alternatively, a qualitatively similar response could come from a non-homogeneous oxide thickness across the film that varies from 1 to 2 SL, and, thereby, leads to portions of the film which are effectively an even number of layers and portions of the film with an effective odd number of layers. For example, if the films were effectively 5/6 SL or 4/5 SL the magnitude $R_{xy,F}$ would likely be significantly smaller, which makes the scenario of the magnetic split across thickness perhaps more likely. Overall, these data and analysis clearly show an unexpected response of the anomalous Hall components. We have interpreted these data in the context of the model of Ref. [48], yet fully recognizing that the rich anomalous Hall response in MnBi$_2$Te$_4$ will require additional detail from theoretical and experimental studies to fully understand, for example, the origin of the non-compensated components in even layer samples, as well as the role of growth- and interface-related defects in regards to the impurity-like term.

To begin to comprehend how both the anomalous Hall effect and quantum anomalous Hall effect arise in this material requires understanding the role and effect of oxide formation on the band structure as well as the magnetism. To address these effects in the context of our experimental results, first principles calculations were performed to show the electronic and magnetic structures with 1 ML of oxygen atoms within the SL structure (see Figure 1(d) for the optimized structure), which are compared with those of the O-free system. The formation energy of 1 ML of O$_i$ is calculated to be -0.592 eV per O$_i$. Figure 6 shows several calculations: First, the electronic band structure of bulk MnBi$_2$Te$_4$ with the A-type AFM ground state calculated and is shown in Figure 6(a), which is extended to a slab calculation of a 7 SL film shown in Figure 6(b). Here, the band gap is around 0.21 eV for the bulk case, and, for the 7 SL slab within the bulk band gap there is a topological surface state that is gapped because of the net moment. A key difference between the bulk and the 7 SL slab is that the loss of inversion symmetry for the 7 SL calculation introduces weak lifting of the degeneracies, giving rise to additional bands. To understand the effects of oxygen on the top layer, the 7 SL model was extended to include a single monolayer of oxygen. Here the oxygen changes energy scales for the intra-layer Mn-Mn coupling, which favors a checkerboard-like pattern (antiparallel alignment along the orthogonal in-plane directions), which is slightly more stable (by ~0.1 meV per Mn) than the ferromagnetic alignment.

The in-plane AFM alignment can be understood in the context of the Goodenough-Kanamori rules, where the Mn-Te-Mn bond angle is increased to 99.8° in the presence of oxygen relative to the pristine 95.4°. Due to the energetic similarity, however, both orderings are calculated and the results for the ferromagnetic case are shown in Figure 6(c) and the checkerboard case in Figure 6(d). In comparing Figure 6(c-d) with the oxide to the pristine case, Figure 6(b), additional metallic states arise that are spatially confined to the surface region. These states span the exchange gap formed within the topological surface states, and, therefore, should conduct in parallel with a quantized edge mode. Moreover, these states should exhibit a unique anomalous Hall response, which may be necessary to understand the additional channel that emerges at 30-60 minutes, as seen in Figure 5. The key result of these calculations is that the band structures show the emergence of additional in-gap electronic bands and that multiple surface magnetic states are close in energy. The calculations imply that additional metallic states will arise near the surface, and the magnetism will be altered which will rebalance the net magnetic ground state of the film. It is critical to note that these calculations show an ideal case of 1 ML O. The real situation is more complex, which reflects the observed disordered surface and magnetic structures. However, additional near-surface metallic states and an alternate magnetic ground state will arise. This will, in turn, affect the magnetic-field trajectory a sample takes en route to becoming fully magnetized, which is clearly seen in the experimental data.

To understand these calculations in the context of the present set of data and, more broadly, in regards to the QAHE, we examine our results compared to previous studies which prepared exfoliated samples in gloveboxes prior to being transported to cryostats[27–29,52] and films that have utilized capping layers to protect the surface.[48,49] Our XPS measurements show that exposure to rather low oxygen partial pressures already leads to oxidation and that the additional oxidants in air, such as water, can produce more complex oxidation pathways that seem to deviate from those of pure oxygen. Furthermore, the effect for air-exposed samples seems to have a rapid response in transport measurements, showing both short- and long-term evolutions. These samples show an altered non-compensated magnetic state due to finite-thickness that seems to become effectively thinner, possibly splitting into distinct magnetic layers over time, and develop additional metallic states that span the magnetic exchange gap. This thinning was observed via reduction and bifurcation of the spin-flop transition as well as by the change of components of the anomalous Hall effect. The scale of the spin-flop is found to reduce by about 10% from $H_{SF} \approx 3.8$ T to around 3.5 T across the timescale used here, as shown in Figure 3. In comparison with thickness-dependent measurements, this reduction of $H_{SF}$ is similar to capped films with 4 SL thickness (3.75 T).[48] This would similarly be consistent with the interpretation of the anomalous Hall effect being an even number of SL at around 10 minutes (i.e. the 6 SL physical thickness versus the effective electronic or magnetic thickness), which evolved to an odd number of SL at longer times.

The most significant result of this study is that the "effective thickness" corresponding to the $MnBi_2Te_4$ magnetic and electrical properties is different from that expected for the actual thickness of the films. This realization is particularly important for exploiting the functionality of $MnBi_2Te_4$ in practical applications and requires better understanding of the mechanisms that govern the behavior of finite-thickness thin films and flakes of $MnBi_2Te_4$. This disparity is studied here for a surface oxide, but it will likely be present with other structural and compositional surface perturbations, growth-related defects, or even low levels of surface damage accompanying exfoliation of flakes.

In addition to environmental exposure, device fabrication processing steps can introduce extra perturbation for both thin films and exfoliated $MnBi_2Te_4$ flakes. Specifically, our NAP XPS measurements show that a low background of $P_{O2}$ is sufficient to oxidize the surface over the course of hours, and the ex-situ XPS measurements suggest exposure to other oxidants like water rapidly accelerate the oxidation process. For example, most studies report that care must be taken to maintain pristine surfaces, for example, by electrode patterning using chemical-free stencil masks.[27–29,52] However, the lone observation of the QAHE with quantization being stable at zero field,[27] utilized $Al_2O_3$-films deposited on $MnBi_2Te_4$ crystals to aid exfoliation, suggesting that oxidation related perturbations of the surface cannot be ruled out. These results indicate that the effects of extended exposure to small amounts of oxygen/water and chemical processing (photoresists, developers, etchants, either physical or chemical, and thermal processing steps) as well as physical surface damage during exfoliation needs to be explicitly considered in formation or modification of the surface layer, since these changes can have a dramatic effect on the intrinsic properties. These considerations are directly connected to the replication challenges for achieving samples hosting the QAHE and demonstrate that small perturbations in the chemistry of samples lead to dramatic shifts in the electronic and magnetic structure of the system. This is technically significant knowledge critical to design, observe, and utilize the QAHE at elevated temperatures in functional devices.

## 3. Conclusion

In conclusion, we present a combination of spectroscopy, diffraction, electrical and magnetic transport data, with theoretical calculations to show unambiguously that a surface oxide layer begins to form within minutes of air exposure of the magnetic topological insulator $MnBi_2Te_4$ thin films. The formation of the surface oxide is found to be kinetically limited to the top one or two SL layers. It alters the properties of $MnBi_2Te_4$ thin films, which is reflected in a complex behavior of transport data. This includes an effective reduction and bifurcation of magnetic and electronic thickness, which drives a reversal of the anomalous Hall response. These findings represent intriguing puzzles regarding what specifically determines the magnetic and electronic thicknesses in these materials. Our work identifies critical aspects that can illuminate current questions in the literature regarding the conditions required to stabilize the quantum anomalous Hall effect in MnBi$_2$Te$_4$, and overall reproducibility of this phenomenon. The data reveal the delicate nature of the magnetic and topological properties in this material system, which opens many new routes for tailoring properties for novel functionalities. Specifically, it is shown that slight perturbation of the surface layers can dramatically rebalance the magnetic energy scales and enable different electronic responses of the anomalous Hall effect. These results open the possibility for tuning quantum phenomena by selectively functionalizing thin films—a prospect which promises exciting potentials for real-world applications.

## Data Availability
The data that support the findings of this study are available from the corresponding author upon reasonable request.

## Supporting Information
See Supporting Information for additional X-ray measurements and details for the first principles calculations and neutron reflectometry. This includes references [45,53–63].

## Acknowledgements

This work was supported by the U. S. Department of Energy (DOE), Office of Science, Basic Energy Sciences (BES), Materials Sciences and Engineering Division (growth, neutron, transport, electron microscopy, spectroscopy and density functional calculations), and the National Quantum Information Science Research Centers, Quantum Science Center (structure). A portion of this research used resources at the Spallation Neutron Source, a DOE Office of Science User Facility operated by the Oak Ridge National Laboratory. The author's wish to thank Jiaqiang Yan and Andrew F. May for insightful discussions and Ryan Comes and Patrick Gemperline for insight into the XPS measurements.

# Figures

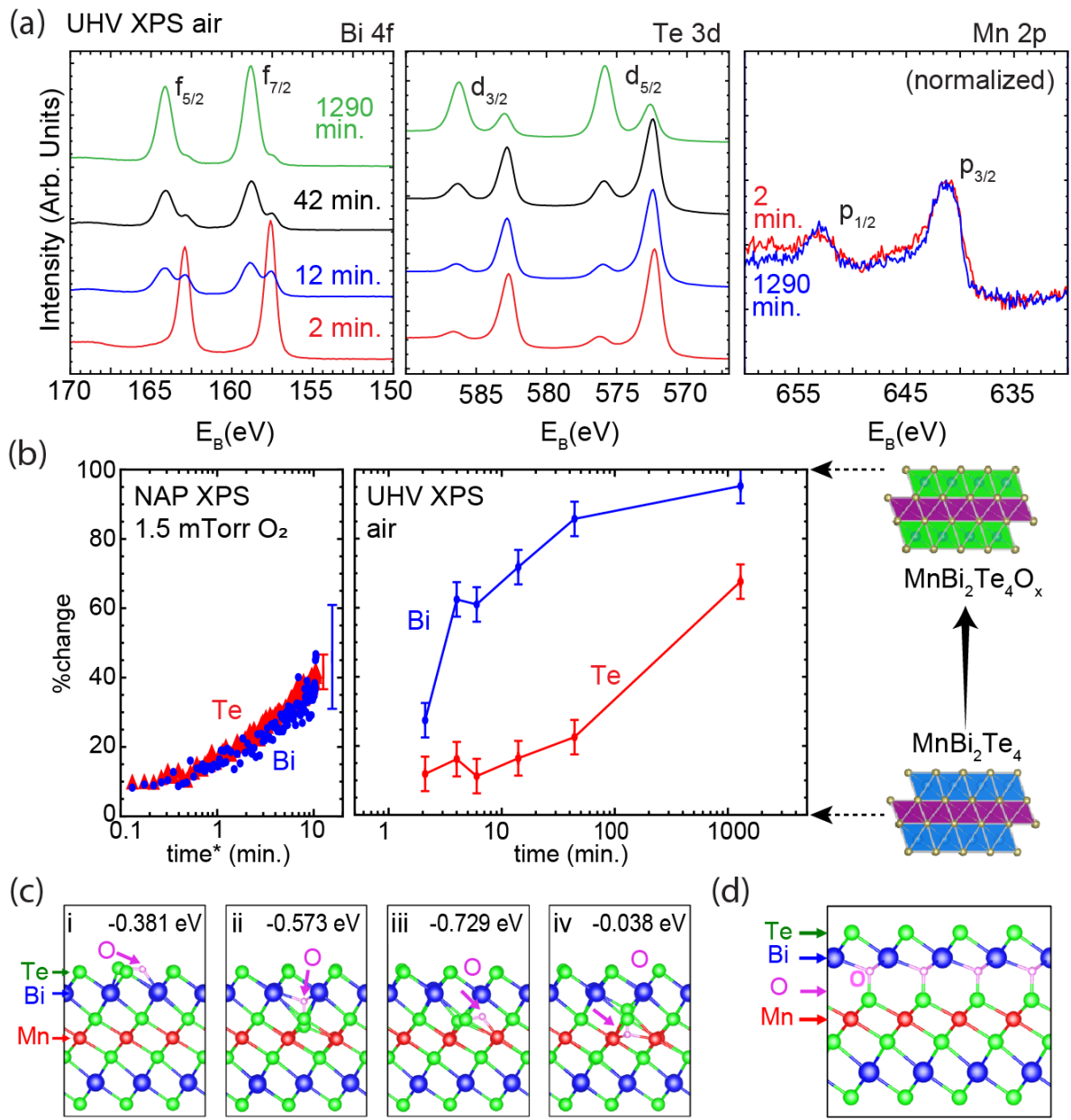

**Figure 1.** (a) XPS spectra for Bi 4f (left), Te 3d (middle), and Mn 2p (right), plotted as a function of exposure times in air. Mn spectra are plotted at the endpoints to show the lack of change with air exposure. See Figure S2 for fits. (b) Percentage of the total integrated intensity (sum of oxide and metal peaks) of the oxide peak in the XPS spectra for Bi and Te versus exposure time. The left panel shows data from real-time measurements in $P_{O2,XPS} \approx 1.5$ Torr of oxygen ($time^* = P_{O2,XPS}/P_{O2,air} \times time$) and the right panel shows data from air exposed samples for the evolution of oxide layers. (c) Results of first principles calculations which schematically show the formation energies for an oxygen atom within the MnBi₂Te₄ septuple layer structure. (d) Schematic showing a relaxed fully occupied ML of oxygen bonded with about a ≈110° Te-O-Bi bond angle in the SL structure.

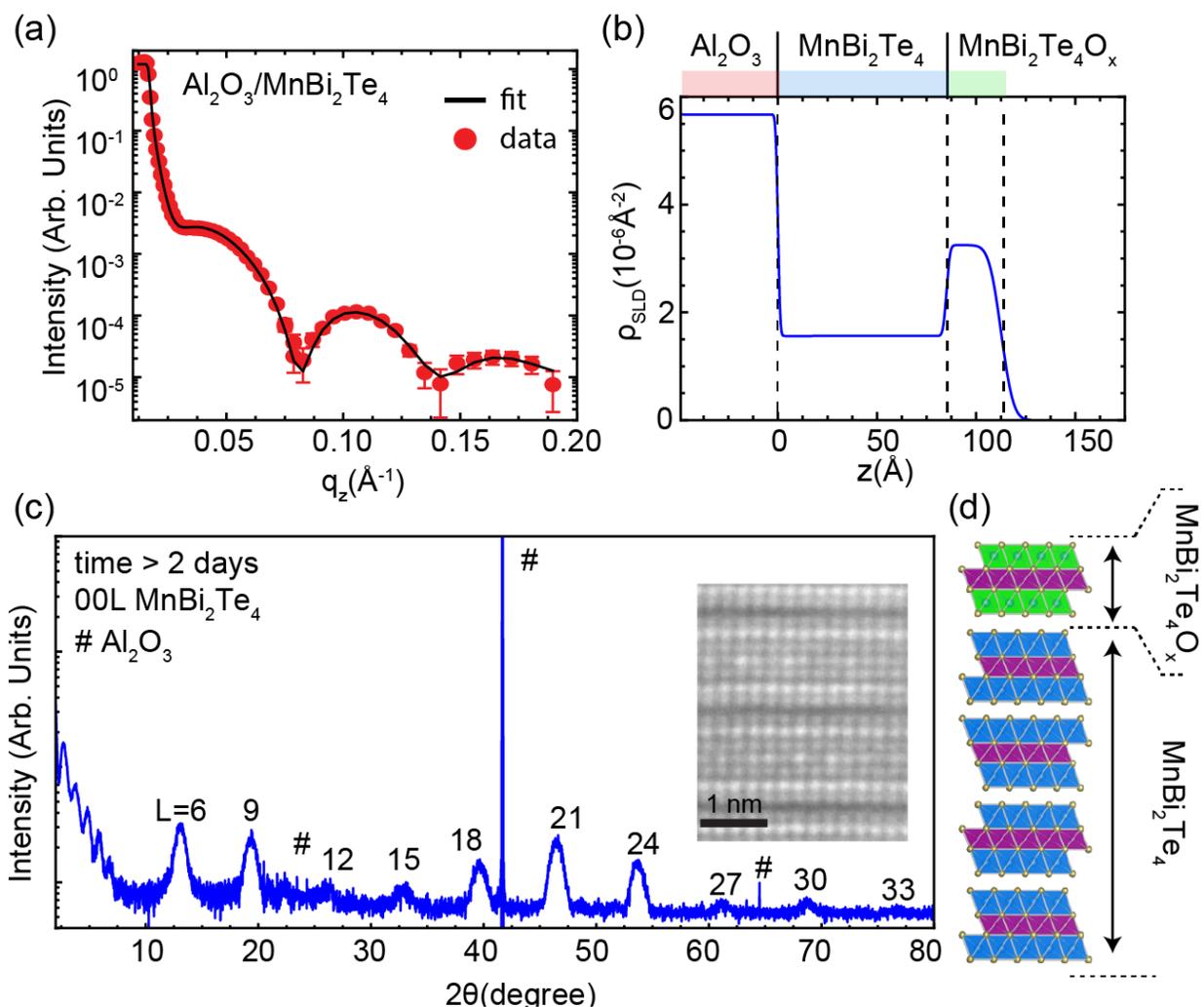

**Figure 2.** (a) Neutron reflectivity results for a MnBi$_2$Te$_4$ film. The solid line is the best fit to the data described in the text and Supporting Information. (b) Scattering length density profile, $\rho_{SLD}$, corresponding to the fit in (a), showing a clear oxide formation of ~1-2 nm thickness in MnBi$_2$Te$_4$ as indicated by a green bar at the top as MnBi$_2$Te$_4$O$_x$. (c) X-ray diffraction 2θ scan showing well defined 00L MnBi$_2$Te$_4$ peaks, indicating that the bulk structure of the film is preserved. The inset shows the SL structure of the MnBi$_2$Te$_4$ film using STEM. (d) Schematic depiction of the oxide on the surface layer and the underlying pristine bulk.

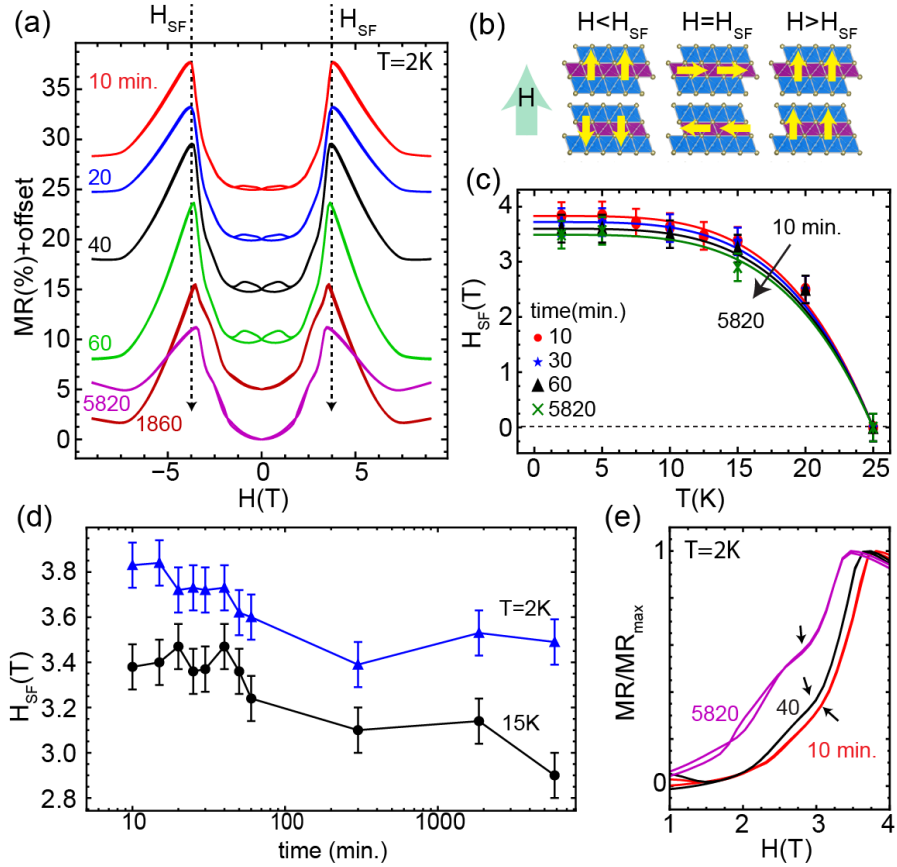

**Figure 3.** (a) Magnetoresistance, MR, as a function of magnetic field. Results measured at various air exposure times indicated left (curves are offset by 5%), show changes with increasing oxidation. These measurements are performed at a temperature of 2 K, and the dashed vertical arrows indicate the spin-flop transition field, $H_{SF}$. (b) Schematic illustration of the magnetic-field-induced spin transitions. Spin-flop field versus temperature at various times (c) and as a function of time (d) for temperature of 2 and 15 K. (e) Detailed view of the data in (a) normalized to the maximum MR.

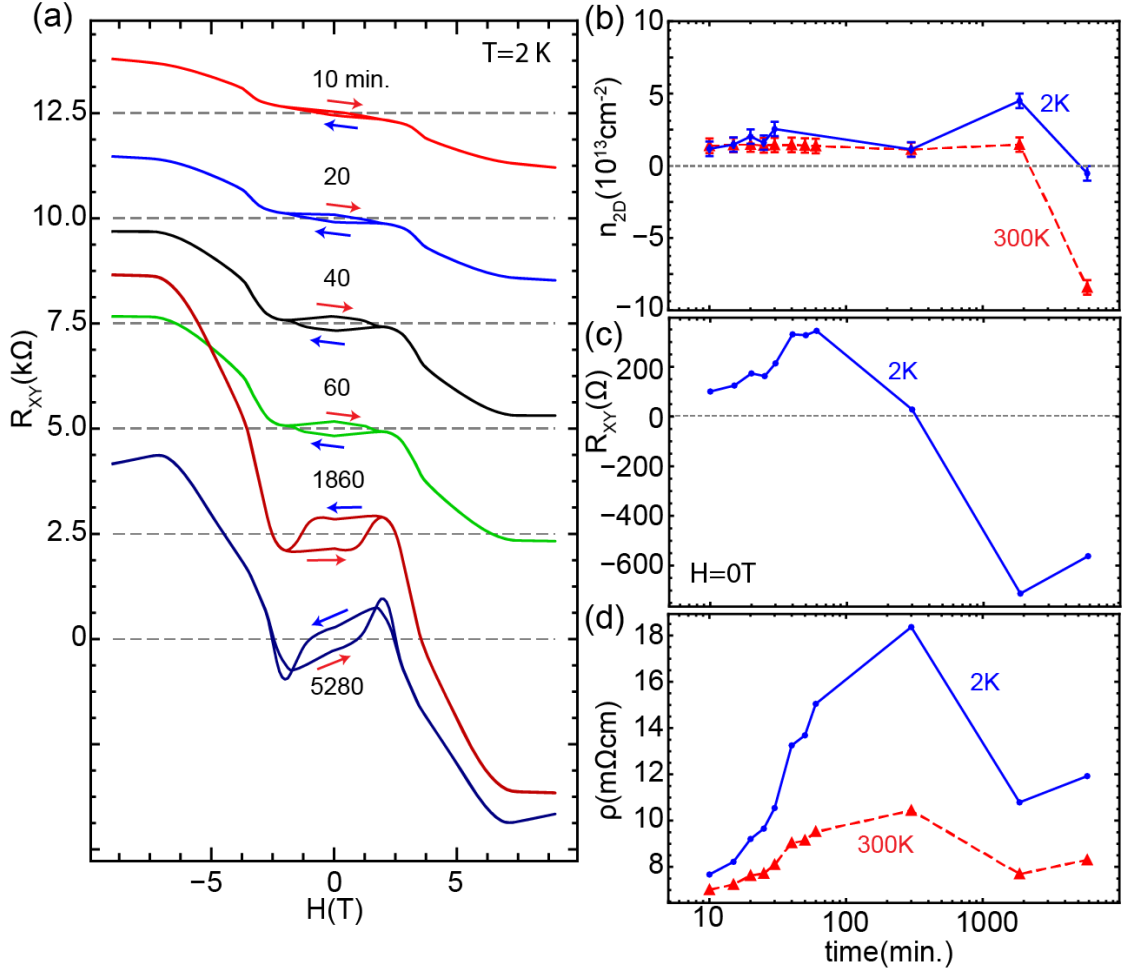

**Figure 4.** (a) Hall effect measurements at 2 K for various exposure times (curves are offset by 2.5 kΩ). (b) Carrier density versus time for temperatures of 2 K (blue) and 300 K (red). These data are extracted from the slope of $R_{xy}$ in the high field regime for 2 K (a), and across the full field range for 300 K. (c) $R_{xy}$ versus exposure time. Data are taken at zero magnetic field from (a) on the up-sweep. (d) Resistivity versus exposure time at temperatures of 2 K and 300 K.

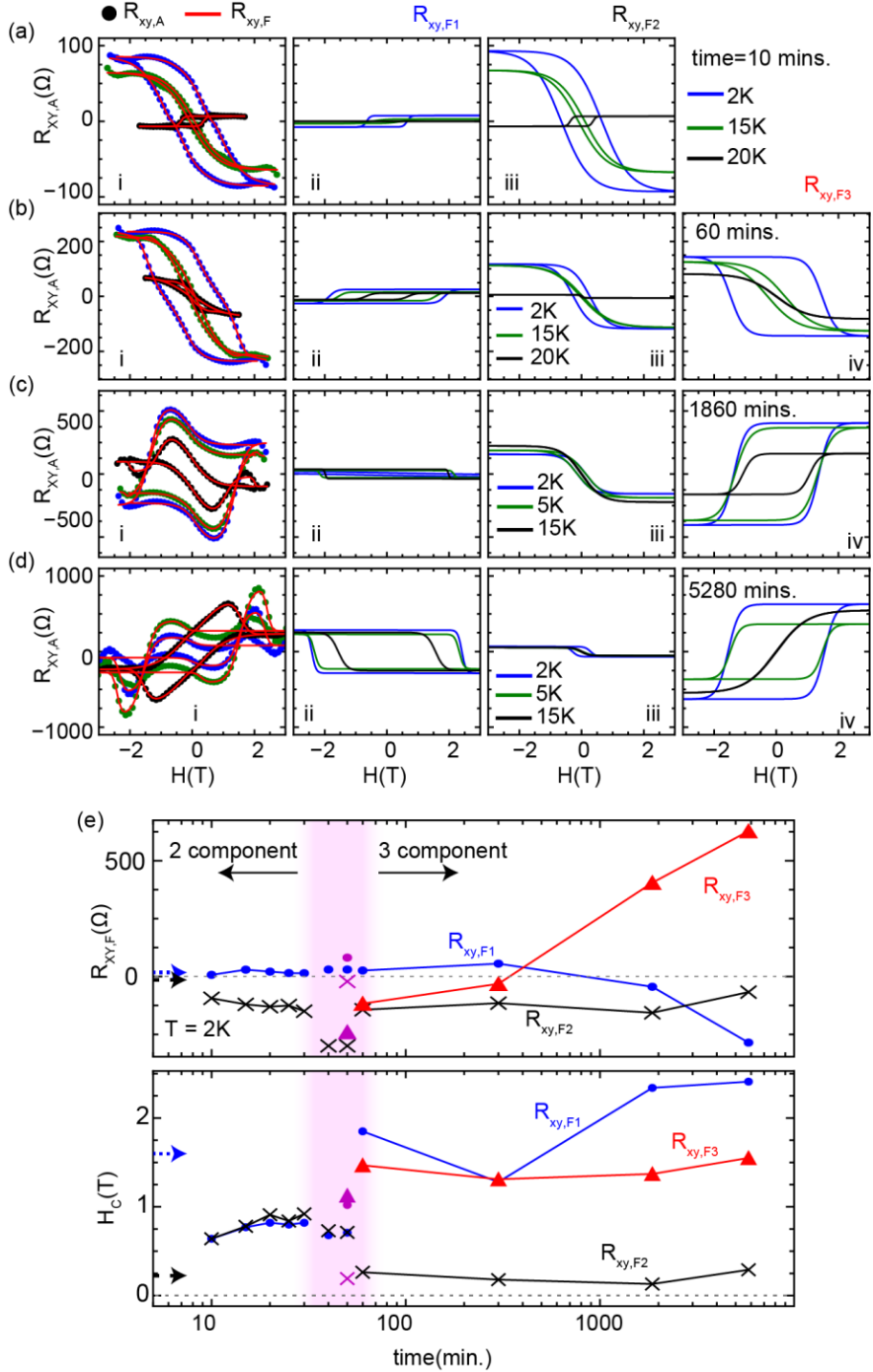

**Figure 5**. (a-d) Fitting results for the hysteretic portion of the Hall effect (full temperature range shown in Figure 4(a)) with exposure time and temperatures indicated on the right panel. Column i shows data (symbols) and fit (solid lines). Columns ii-iv depict the individual anomalous Hall components. As shown, the data can be seen as splitting into two components, where the F2 component (column iii) is no longer sufficient to describe the data and a third component, F3 (column iv), is required to fit the data after 60 minutes. (e) Fitted components plotted as a function of time. The arrows in this panel are extracted from Ref. [48] to represent data taken prior to air exposure. The top panel shows $R_{xy,F}$ and bottom panel shows the respective $H_C$.

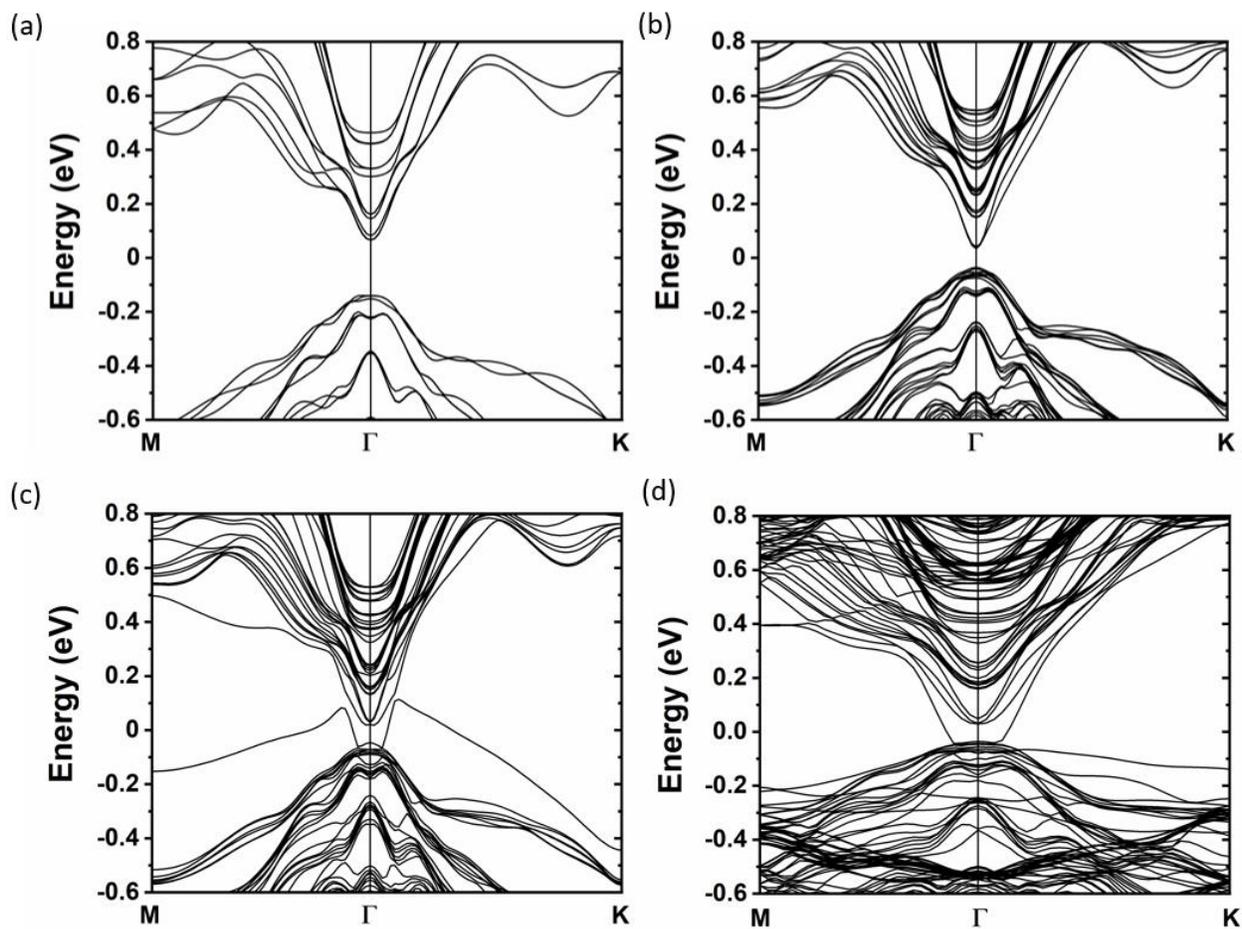

**Figure 6.** (a-d) Electronic band structures of MnBi$_2$Te$_4$ plotted in the surface Brillouin zone for bulk (a), for a MnBi$_2$Te$_4$ thin film consisting of 7 SL (b), 7 SL with 1 ML of O$_i$ with ferromagnetic (c) and antiferromagnet (d) intra layer ordering of the Mn-moments within the top layer. Note that a 2×2 surface cell was used for the band structure calculation in (d) to allow the antiferromagnet intra-layer coupling, resulting in a smaller Brillouin zone.

# Supplemental Information
## for
**Surface-Driven Evolution of the Anomalous Hall Effect in Magnetic Topological Insulator MnBi$_2$Te$_4$ Thin Films**

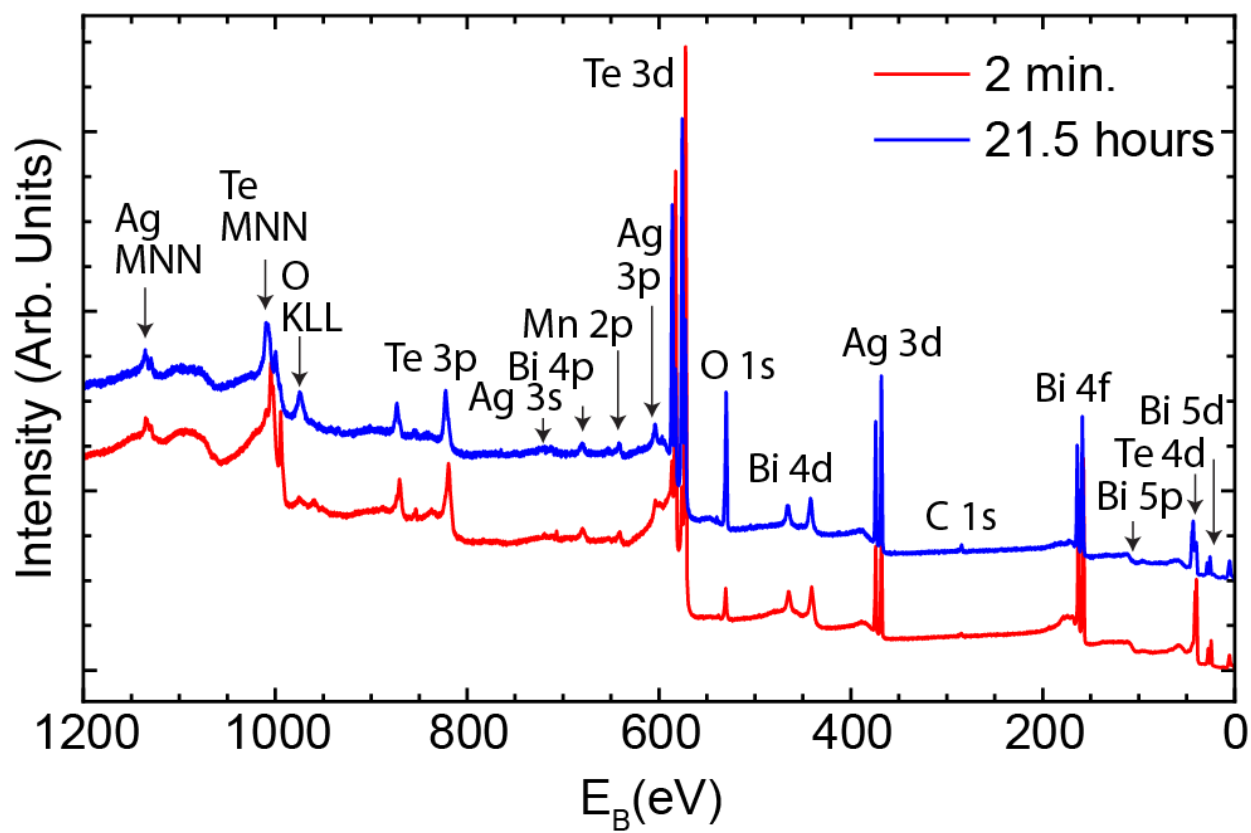

**Figure S1** XPS survey spectra of the MnBi$_2$Te$_4$ film after 2 minutes (red line) and 21.5 hours (blue line) exposure to air. Ag is present from Ag paste used in mounting the sample.

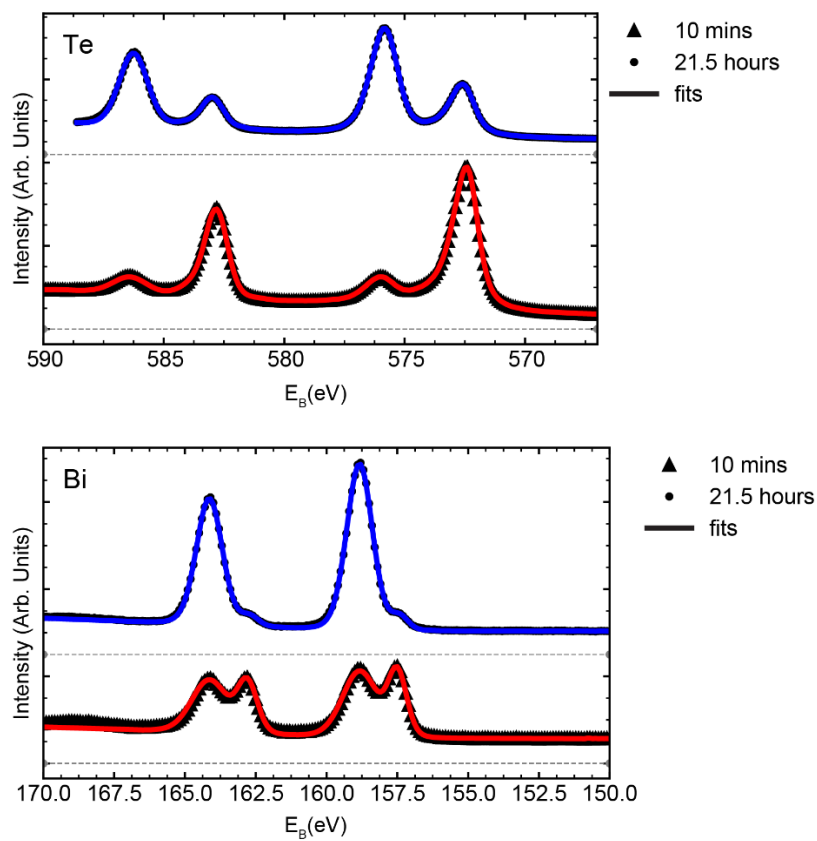

**Figure S2** Sample XPS data (symbols) and fits (solid lines) for MnBi$_2$Te$_4$ films shown in Figure 1(a) of the main text, which was used to determine the ratio of the oxide and metal peaks in Figure 1(b).

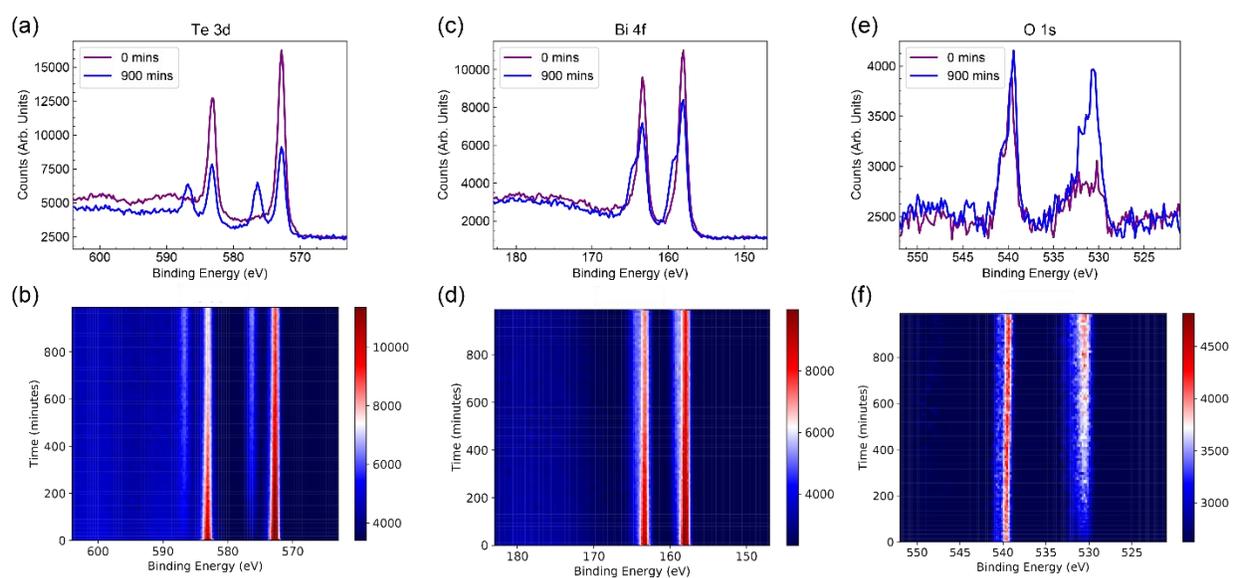

**Figure S3** *Real-time* environmental XPS for MnBi$_2$Te$_4$ films exposed to 1.5 Torr $P_{O2}$. (a) shows the Te 3d spectra at initial and final time with the full evolution being shown in (b). Similar data for the Bi 4f (c-d) and the O 1s (e-f), which show a similar trend in the formation of a surface oxide in this low oxygen environment.

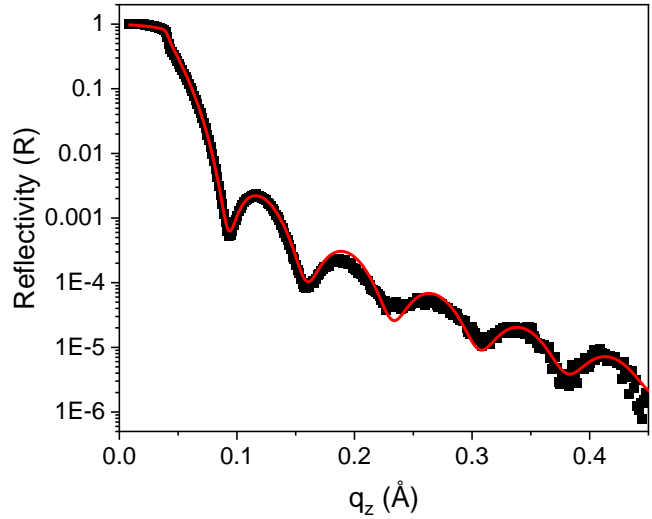

**Figure S4** X-ray reflectivity (XRR) data (black dots) used to determine the thickness of the sample studied as a function of air exposure in transport measurements. From the fit (in red) shown it was found that the sample thickness is 83.4 ± 0.5 Å or nominally 6.1 septuple layers (SL) using the c-axis lattice parameter of 13.65 Å, as extracted from the fits in Fig. S3.

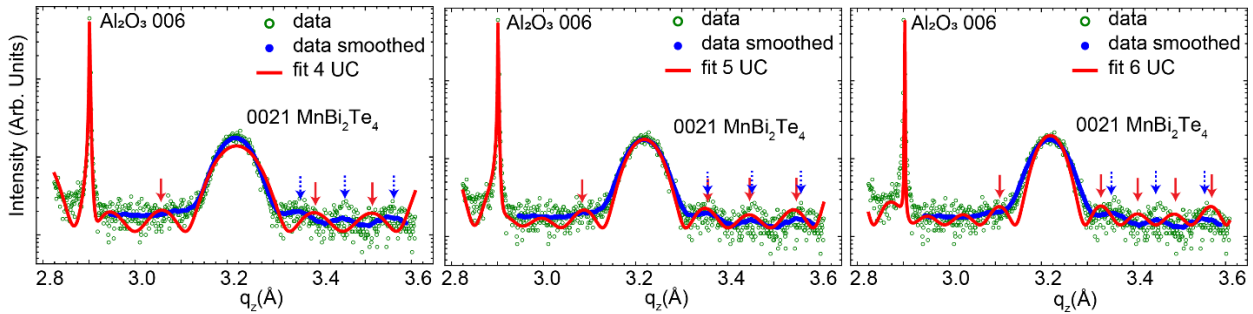

**Figure S5** Data and fits to x-ray diffraction from the film measured in transport after air exposure (Figs. 3-5 main text). Green symbols are the data, blue are the smoothed data (±20 data points), and solid line is the fit with thickness fixed at 4 (left), 5 (middle), and 6 SL (right). See Refs. [1,2] for details on the fitting. In contrast to the physical thickness from XRR (~6.1 SL) the crystalline MnBi$_2$Te$_4$ thickness is shown best fit to be 5 SL, which can be seen by comparing the positions of the maxima of oscillation, as highlighted by the arrows. This is consistent with the change seen in transport where the character changes from 6 SL to 5 SL in nature as a function of air exposure.

**Density functional theory methods**

All calculations are based on density functional theory (DFT)[3,4] implemented in the VASP code.[5] The interaction between ions and electrons is described by the projector augmented wave method.[6] The total energy is calculated using the Perdew-Burke-Eznerhof (PBE) exchange correlation functional[7] and a kinetic energy cutoff of 400 eV. A $U$ parameter of 3 eV is applied to Mn 3d orbitals[8] and the DFT-D3 vdW functional[9] is used, following several previous DFT studies. A 3×3 surface cell was used in calculations of isolated O interstitials in the $MnBi_2Te_4$ thin film consisting of 7 SLs. A 1×1 or 2×2 surface cell was used in the case of 1 ML of O interstitials. Lattice parameters were optimized, and atomic positions were relaxed until the forces are less than 0.01 eV/Å. The optimized lattice constants of $MnBi_2Te_4$ are $a$ = 4.355 Å and $c$ = 40.596 Å, which are in excellent agreement with the experimental values of $a$ = 4.334 Å and $c$ = 40.931 Å[10].

The formation energy of a neutral O interstitial is given by

$$\Delta H(O_i) = E_D - E_h - \mu_O, \tag{1}$$

where $E_D$ and $E_h$ are the total energies of the oxygen-containing and the host (i.e. oxygen-free) supercells, and $\mu_O$ is the O chemical potential. The O chemical potential at the ambient condition [$T$ = 300 K and $P(O_2)$ = 0.21 atm] was calculated following the method in Ref.[11] We calculated $\Delta H(O_i)$ for 1 ML of O interstitial with and without the spin-orbit coupling (SOC); the difference was found to be only 0.2 meV per $O_i$. Therefore, we calculated $\Delta H(O_i)$ for other systems without the spin-orbit coupling. However, the energy difference between the FM and AFM Mn-Mn intralayer coupling was calculated including the SOC because such energy difference is very small and electronic band structures were all calculated including SOC.